\documentclass[12pt]{iopart}
\usepackage{graphicx}
\usepackage{amssymb}
\usepackage{bm}
\begin{document}
\title[Specific heats of Ne inside single-walled carbon nanotube]{Specific heats of dilute neon inside long single-walled carbon nanotube}
\author{Z. C. Tu\dag\ddag\footnote[7]{To
whom correspondence should be addressed (tzc@itp.ac.cn)} and Z. C.
Ou-Yang\dag\S}
\address{\dag\ Institute of
Theoretical Physics,
 The Chinese Academy of Sciences,
 P.O.Box 2735 Beijing 100080, China}
 \address{\ddag\ Graduate School,
 The Chinese Academy of Sciences, Beijing, China}
 \address{\S\ Center for Advanced Study,
 Tsinghua University, Beijing 100084, China}
\begin{abstract} An elegant formula for coordinates of
carbon atoms in a unit cell of a single-walled nanotube (SWNT) is
presented and the potential of neon (Ne) inside an infinitely long
SWNT is analytically derived out under the condition of the
Lennard-Jones potential between Ne and carbon atoms.
 Specific heats of dilute Ne inside long (20, 20) SWNT are calculated at different
temperatures. It is found that Ne exhibits 3-dimensional (3D) gas
behavior at high temperature but behaves as 2D gas at low
temperature. Especially, at ultra low temperature, Ne inside (20,
20) nanotubes behaves as lattice gas. A coarse method to determine the
characteristic temperature $\mathcal{T}_c$ for low density gas in a potential is put
forward. If $\mathcal{T}\gg \mathcal{T}_c$, we just need to use
the classical statistical mechanics without solving the
Shr\"{o}dinger equation to consider the thermal behavior of gas in
the potential. But if $\mathcal{T}\sim \mathcal{T}_c$, we must
solve the Shr\"{o}dinger equation. For Ne in (20,20) nanotube, we
obtain $\mathcal{T}_c\approx 60$ K.
\end{abstract}
\submitto{\JPCM}  \maketitle

\section{\label{intro}Introduction}
Since the discovery of carbon nanotubes \cite{s}, the peculiar
electronic and mechanical properties of these structures have
attracted much attention \cite{jw,bi,x}. Experiments have also
revealed that they can also be used to store hydrogen
\cite{dillon} and other gases \cite{teizer}. Many physicists
expected that gases in nanotubes or nanotube bundles may display
novel 1-dimensional (1D) behavior as a consequence of the
remarkable aspect ratio of the length of tubes to their radius.
The group led by Cole \cite{cole} and other researchers
\cite{carraro} have theoretically studied properties of gases in
nanotubes or nanotube bundles. One of the most fantastic
properties they found is specific heat of dilute gas inside
single-walled carbon nanotubes (SWNT's) as a functions of
temperature: With temperature increasing it shows the thermal
behavior changing from 1D, 2D, and 3D. However, there are still
two question arisen: Do 1D and 2D behaviors always exist for
dilute gas inside SWNT only if the temperature is low enough? Is
classical statistical mechanics (CSM) sufficient to deal with this
problem?

In Ref. \cite{tzc1}, we has given a brief answer to the first
question. Here we will exhibit the full calculations and continue
to discuss the second question. In recent papers, \u{S}iber {\it et al}
treat the specific heat of dilute He atoms adsorbed in interstitial
channels and grooves of carbon nanotube bundles \cite{siber}. Here we partially follow their aprroach.

This paper is organized as follows: In Sec.\ref{coord}, we give
the expressions of the coordinates of carbon atoms in a unit cell
of a SWNT. In Sec.\ref{potential}, we analytically calculate the
potential of Ne in a carbon nanotube. For example, we calculate
the potential of Ne in (20,20) tube. In Sec.\ref{statis}, we list
the related formula of thermodynamics in statistical mechanics to
calculate the specific heat. In Sec.\ref{cvhigh}, we calculate the
specific heats of Ne inside (20,20) at different temperatures
without considering $\theta,z$ effects ($\theta,z$ are defined in Sec.\ref{coord}). In Sec.\ref{cvlow}, we
calculate the specific heats of Ne inside (20,20) at low
temperature. In Sec.\ref{classic}, we discuss why we do not use
CSM to calculate the specific heat. In this section, we present a
coarse method to estimate the characteristic temperature, below
which the CSM can not be used. In
Sec.\ref{discuss}, we discuss the reliability of our results and
give brief conclusions.

\section{\label{coord}the coordinates of carbon atoms in a unit cell of a SWNT}
\begin{figure}[!htp]
\begin{center}\includegraphics[width=7cm]{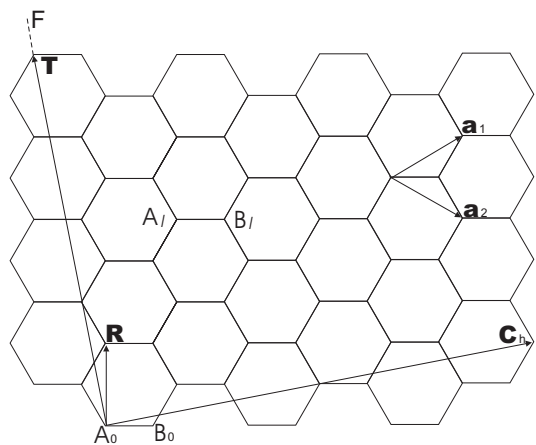}\end{center}
\caption{\label{fig1}The unrolled honeycomb lattice of a SWNT. By
rolling up the sheet along the chiral vector ${\bf C}_h$, that is,
such that the point $A_0$ coincides with the point corresponding
to vector ${\bf C}_h$, a nanotube is formed. The vectors ${\bf
a}_{1}$ and ${\bf a}_{2}$ are the real space unit vectors of the
hexagonal lattice. The translational vector ${\bf T}$ is
perpendicular to ${\bf C}_h$ and runs in the direction of the tube
axis. The vector ${\bf R}$ is the symmetry vector. $A_0$, $B_0$
and $A_l, B_l (l=1,2,\cdots,N)$ are used to denote the sites of
carbon atoms.}
\end{figure}

A SWNT without two caps can be constructed by wrapping up a single
sheet of graphite such that two equivalent sites of hexagonal
lattice coincide \cite{r1}. To describe the SWNT, some
characteristic vectors require introducing. As shown in
Fig.\ref{fig1}, the chiral vector ${\bf C}_{h}$, which defines the
relative location of two sites, is specified by a pair of integers
$(n_1, n_2)$ which is called the index of the SWNT and relates
${\bf C}_{h}$ to two unit vectors ${\bf a}_{1}$ and ${\bf a}_{2}$
of graphite (${\bf C}_{h}=n_1{\bf a}_{1}+n_2{\bf a}_{2}$). The
chiral angle $\theta_0$ defines the angle between $\mathbf{a}_1$
and $\mathbf{C}_h$. For $(n_1, n_2)$ nanotube,
$\theta_0=\arccos[\frac{2n_1+n_2}{2\sqrt{n_1^2+n_2^2+n_1n_2}}]$.
The translational vector ${\bf T}$ corresponds to the first
lattice point of 2D graphitic sheet through which the line normal
to the chiral vector ${\bf C}_{h}$ passes. The unit cell of the
SWNT is the rectangle defined by vectors ${\bf C}_{h}$ and ${\bf
T}$, while vectors ${\bf a}_{1}$ and ${\bf a}_{2}$ define the area
of the unit cell of 2D graphite. The number $N$ of hexagons per
unit cell of SWNT is obtained as a function of $n_1$ and $n_2$ as
$N=2(n_1^2+n_2^2+n_1n_2)/d_R$, where $d_R$ is the greatest common
divisor of ($2n_2+n_1$) and ($2n_1+n_2$). There are $2N$ carbon
atoms in each unit cell of SWNT because every hexagon contains two
atoms. To denote the $2N$ atoms, we use a symmetry vector ${\bf
R}$ to generate coordinates of carbon atoms in the nanotube and is
defined as the site vector having the smallest component in the
direction of ${\bf C}_h$. From a geometric standpoint, vector
${\bf R}$ consists of a rotation around the nanotube axis by an
angle $\Psi=2\pi/N$ combined with a translation $\tau$ in the
direction of ${\bf T}$; therefore, ${\bf R}$ can be denoted by
${\bf R}=(\Psi|\tau)$. Using the symmetry vector ${\bf R}$, we can
divide the $2N$ carbon atoms in the unit cell of SWNT into two
classes: one includes $N$ atoms whose site vectors satisfy
\begin{equation}\label{sitea}
{\bf A}_l=l{\bf R}-[l{\bf R}\cdot{\bf T}/{\bf T}^2]{\bf T} \quad
(l=0,1,2,\cdots,N-1),\end{equation} another includes the remainder
$N$ atoms whose site vectors satisfy
\begin{eqnarray}\label{siteb}&&{\bf B}_l=l{\bf R}+{\bf
B}_0-[(l{\bf R}+{\bf B}_0)\cdot{\bf T}/{\bf T}^2]{\bf T}\nonumber
\\&&-[(l{\bf R}+{\bf B}_0)\cdot{\bf C}_h/{\bf C}_h^2]{\bf C}_h
\quad (l=0,1,\cdots,N-1),\end{eqnarray}
 where ${\bf B}_0\equiv\left (\Psi_0|\tau_0\right)=\left (\left.\frac{2\pi
a_{cc}\cos(\theta_0-\frac{\pi}{6})}{|\mathbf{C}_h|}\right|a_{cc}\cos(\theta_0-\frac{\pi}{6})\right)$
represents one of the nearest neighbor atoms to {\bf A}$_0$.

If we introduce a cylindrical coordinate system $(r, \theta, z)$
whose $z$-axis is the tube axis. Its $r\theta$-plane is
perpendicular to $z$-axis and contains atom $A_0$ in the nanotube.
$r$ the distance from some point to $z$-axis, and $\theta$ the
angle rotating around $z$-axis from an axis which is vertical to
$z$-axis and passes through atom $A_0$ in the tube to the point.
In this coordinate system, we can express Eqs.(\ref{sitea}) and
(\ref{siteb}) as
\begin{equation}\label{siteaa}
{\bf A}_l=(\rho,l\Psi,l\tau-[l\tau/T]T)\quad (l=0,1,2,\cdots,N-1)
,\end{equation} and
\begin{eqnarray}\label{sitebb} {\bf
B}_l&=&(\rho,l\Psi+\Psi_0-2\pi[\frac{l\Psi+\Psi_0}{2\pi}],l\tau+\tau_0-[\frac{l\tau+\tau_0}{T}]T)\nonumber\\
&&(l=0,1,2,\cdots,N-1),\end{eqnarray} where
$\rho=\frac{|\mathbf{C}_h|}{2\pi}$. In
Eqs.(\ref{sitea})-(\ref{sitebb}), the symbol $[\cdots]$ denotes the largest integer smaller than $\cdots$, e.g., $[5.3]=5$.

\section{\label{potential}the potential of ne inside carbon nanotubes}
\begin{figure}[!htp]
\begin{center}\includegraphics[width=7cm]{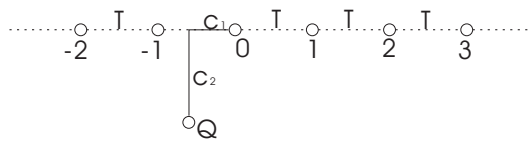}\end{center}
\caption{\label{fig2}An infinite atom chain and an atom Q out of
the chain. Many atoms distribute regularly in a line form the
infinite atom chain. The interval between neighbor atoms in the
chain is $T$, and the site of atom Q relative to atom 0 can be
represented by numbers $c_1$ and $c_2$.}
\end{figure}

To obtain the potential of Ne inside the nanotube, we firstly
consider another simple system shown in Fig.~\ref{fig2}: Many
atoms distribute regularly in a line form an infinite atom chain
and an atom Q is out of the chain. The interval between neighbor
atoms in the chain is $T$, and the site of atom Q relative to atom
0 can be represented by numbers $c_1$ and $c_2$. We take the
Lennard-Jones potential
$U(R_j)=4\epsilon[(\sigma/R_j)^{12}-(\sigma/R_j)^6]$ between atom
Q and atom $j$ in the chain, where $R_j$ is the distance between Q
and atom $j$, and $\epsilon=\sqrt{\epsilon_c\epsilon_{ne}}$,
$\sigma=(\sigma_c+\sigma_{ne})/2$ with $\epsilon_{ne}=35.6$ K,
$\sigma_{ne}=2.75$ \AA,\ \ $\epsilon_c=28$ K and $\sigma_c=3.4$
\AA \cite{cole,hir}. We calculate the potential between atom Q and
the chain as
\begin{equation} \label{upc}
U_{QC}=4\epsilon[\sigma^{12}U_6(c_1,c_2)-\sigma^6U_3(c_1,c_2)]
,\end{equation}where
$U_k(c_1,c_2)=\sum\limits_{n=-\infty}^{\infty}\frac{1}{[(c_1+nT)^2+c_2^2]^k}\quad
(k=1,2,\cdots)$ which can be calculated through the following
recursion \cite{anran}:
\begin{equation}\label{uk}
\begin{array}{l} U_1(c_1,c_2)=\frac{\pi
\sinh (2\pi c_2/T)}{c_2T[\cosh (2\pi c_2/T)-\cos (2\pi c_1/T)]},\\
U_{k+1}(c_1,c_2)=-1/(2kc_2) \partial U_k/\partial c_2. \end{array}
\end{equation}

\begin{figure}[!htp]
\begin{center}\includegraphics[width=7cm]{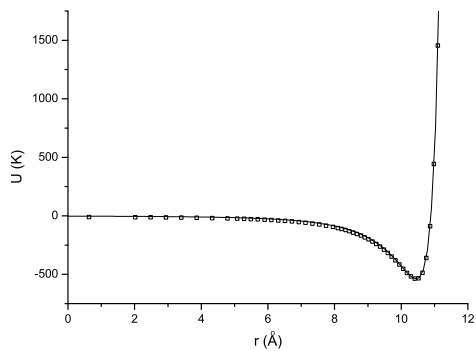}\end{center}
\caption{\label{fig3} The potentials inside the $(20,20)$ nanotube
calculated from Eq.(\ref{potential}) (squires) and the fit curve
(solid curve).}
\end{figure}

The $(20, 20)$ tube, for example, with infinite length can be
regarded as $2N=80$ chains. Thus the potential of any point Q
inside the tube can be calculated as
\begin{equation}\label{potential}U(r,\theta,z)=\sum_{i=1}^{2N}U_{QC},\end{equation}
where $(r,\theta,z)$ is coordinates of Q in the cylindrical
coordinate system. As an approximation, we neglect the potential
varying with $z$ and $\theta$ because we find that it is much
smaller than the potential varying with $r$ through calculations,
and fit the potential with
$U(r)=4\varepsilon[(\frac{\tilde{\sigma}}{\rho-r})^{10}-(\frac{\tilde{\sigma}}{\rho-r})^{4}]$,
where $\rho=13.56$ \AA\ \ is the radius of the tube,
$\varepsilon=390$ K, and $\tilde{\sigma}=2.63$ \AA \ \ (see also
Fig.\ref{fig3}). Moreover, we simplify it as
\begin{equation}\label{ur}
U(r)=\left
\{\begin{array}{l}4\varepsilon[(\frac{\tilde{\sigma}}{\rho-r})^{10}-(\frac{\tilde{\sigma}}{\rho-r})^{4}]
,\quad r<\rho-\tilde{\sigma},\\
\infty,\quad r>\rho-\tilde{\sigma}.\end{array}\right.
\end{equation}

\section{\label{statis} The basic formulae of thermodynamics in statistical mechanics}
When we deal with the thermal behavior of dilute Ne in nanotube, we will
begin with the free energy $\mathcal{F}=-\mathcal{T}\ln
\mathcal{Z}$, where $\mathcal{T}$ is temperature and
$\mathcal{Z}=\sum_n e^{-E_n/\mathcal{T}}=\mathrm{tr}
\exp(-\mathcal{H}/\mathcal{T})$ is the partition function. We have let the
Boltzmann factor be 1 \cite{landau2}. $E_n$ is the eigenvalues of the
Schr\"{o}dinger equation: $\mathcal{H}\psi_n=E_n\psi_n$. From the
free energy, we can derive out the specific heat
$c_v=-\mathcal{T}(\partial^2 \mathcal{F}/\partial
\mathcal{T}^2)_v=\frac{\langle E_n^2\rangle-\langle
E_n\rangle^2}{\mathcal{T}^2}$, where $\langle E_n^2\rangle=\sum_n
E_n^2 e^{-E_n/\mathcal{T}}/\mathcal{Z}$ and $\langle
E_n\rangle=\sum_n E_n e^{-E_n/\mathcal{T}}/\mathcal{Z}$.

We call above statistical mechanics ordinary statistics (OS) and
list the corresponding formula in CSM: The free energy
$\mathcal{F}_{cl}=-\mathcal{T}\ln \mathcal{Z}_{cl}$, where
$\mathcal{Z}_{cl}=\int'e^{-E(p,q)/\mathcal{T}}d\Gamma$ is the
partition function and here the prime means we integrate only over
the regions of phase space which correspond to physically
different states of particles. The specific heat
$c_{vcl}=-\mathcal{T}(\partial^2 \mathcal{F}_{cl}/\partial
\mathcal{T}^2)_v=\frac{\langle E^2\rangle-\langle
E\rangle^2}{\mathcal{T}^2}$, where $\langle
E^2\rangle=\int'E^2e^{-E(p,q)/\mathcal{T}}d\Gamma/\mathcal{Z}_{cl}$
and $\langle
E\rangle=\int'Ee^{-E(p,q)/\mathcal{T}}d\Gamma/\mathcal{Z}_{cl}$.

Under the sufficient high temperature, $\mathcal{F}$ converge to
the $\mathcal{F}_{cl}$. In other words, the applicable domain of
$\mathcal{F}$ is larger than that of $\mathcal{F}_{cl}$.
Therefore, the conclusion derived from $\mathcal{F}$ is much more
reliable than that derived from $\mathcal{F}_{cl}$. We will use OS
in the following discussions if we do not make special statement.

\section{\label{cvhigh} Specific Heats of Ne gas in an approximate potential}
Because we consider the dilute Ne, we can neglect the interaction
between Ne atoms and write the single particle Schr\"{o}dinger
equation \cite{landau1} as $H\psi=E\psi$, where
$H=-\frac{\hbar^2}{2\mu}\nabla^2+U(r)$ and $\psi=\phi
e^{i(m\theta+\kappa z)}$. It follows that
\begin{equation}\label{schr2}
\label{schr} \begin{array}{l}E=\frac{\hbar^2\kappa^2}{2\mu}+E_m\quad (\kappa\in \mathbb{R}, m=0,\pm 1,\pm 2,\cdots),\\
H(r)\phi=E_m\phi,\\
H(r)=-\frac{\hbar^2}{2\mu}(\frac{d^2}{dr^2}+\frac{1}{r}\frac{d}{dr}-\frac{m^2}{r^2})+U(r),\end{array}
\end{equation}
where $m$ is angular quantum number and $E_m$ is the corresponding energy.
Setting $r=(\rho-\tilde{\sigma})\xi$,
$\varepsilon_0=\frac{\hbar^2}{2\mu\rho^2}$ and
$\eta=\tilde{\sigma}/\rho$, Eqs.(\ref{ur}) and (\ref{schr2}) are
transformed into
\begin{equation}\label{uxi}
u(\xi)=\left
\{\begin{array}{l}4\varepsilon[(\frac{\eta}{1-(1-\eta)\xi})^{10}-(\frac{\eta}{1-(1-\eta)\xi})^{4}]
,\quad \xi<1,\\
\infty,\quad \xi> 1.\end{array}\right.
\end{equation}
and
\begin{equation}\label{schr3}
\label{schr} \begin{array}{l}\mathcal{H}\varphi(\xi)=E_m\varphi(\xi),\\
\mathcal{H}=-\frac{\varepsilon_0}{(1-\eta)^2}(\frac{d^2}{d\xi^2}+\frac{1}{\xi}\frac{d}{d\xi}-\frac{m^2}{\xi^2})+u(\xi).\end{array}
\end{equation}
If we let $|\varphi\rangle=\sum_na_n|\chi_n\rangle$, we will
obtain the secular equation
\begin{equation}\label{scu1}
det(\mathcal{H}_{jn}-E_m\mathcal{S}_{jn})=0,
\end{equation}
where $\mathcal{H}_{jn}=\int_0^1\chi_j\mathcal{H}(\xi)\chi_n\xi
d\xi$, and $\mathcal{S}_{jn}=\int_0^1\chi_j\chi_n\xi d\xi$. If we
let $\chi_n=J_{|m|}(\nu_n\xi)$, where $J_{|m|}(\xi)$ is the m-th
order Bessel function of the first class and $\nu_n$ is the n-th
zero point of Bessel function \cite{wang}, we can calculated
$E_{mn}, (m=0,\pm 1,\pm 2,\cdots; n=1,2,3,\dots)$ from
Eq.(\ref{scu1}).

If there are $\mathcal{N}$ Ne atoms inside the tube, we have the
free energy $\mathcal{F}=-\mathcal{NT}\ln\mathcal{Z}_1$, where
$\mathcal{Z}_1=\sum_{mn}e^{-E_{mn}/\mathcal{T}}\int_{-\infty}^{\infty}
e^{-\frac{\hbar^2\kappa^2}{2\mu\mathcal{T}}}d\kappa$. We can
easily obtain the specific heat per atom is
\begin{equation}
c_v=-\frac{\mathcal{T}\partial^2\mathcal{F}}{\mathcal{N}\partial\mathcal{T}^2}=\frac{1}{2}+\frac{\langle
E_{nm}^2\rangle-\langle E_{nm}\rangle^2}{\mathcal{T}^2},
\end{equation}
where $\langle E_{nm}\rangle=\frac{\sum_{mn}
E_{mn}e^{-E_{mn}/\mathcal{T}}}{\sum_{mn}e^{-E_{mn}/\mathcal{T}}}$
and $\langle E_{nm}^2\rangle=\frac{\sum_{mn}
E^2_{mn}e^{-E_{mn}/\mathcal{T}}}{\sum_{mn}e^{-E_{mn}/\mathcal{T}}}$.

In Fig.\ref{fig4}, the symbols ``$\triangledown$" reflect $c_v$
varying with the temperature $\mathcal{T}$, which implies Ne atoms
inside (20,20) tube behave as 3D gas at high temperature (specific
heat approaches to 3/2) and 2D gas at low temperature (specific
heat is 1). Therefore, we can naturally assume that all atoms in
the valley of potential $U(r)$ at low temperature, i.e. lie on the
shell $S^*$ with radius $\varrho=\rho[1-(5/2)^{1/6}\eta]$.

At high temperature, the Ne atoms maybe evaporate from the nanotube, but our model cannot recover this effect because our potential is infinite near the surface of carbon nanotute.

\section{\label{cvlow} Specific Heats at low temperature}
Now we consider the thermal property of Ne inside the nanotube at
low temperature in detail. We assume that all atoms lie on the
shell $S^*$ with radius $\varrho=\rho[1-(5/2)^{1/6}\eta]$ at temperature lower than some critical temperature $T_{lc}$ which depends on the energy difference between the tow lowest energy of radial excitations. From Eqs.(\ref{upc})-(\ref{potential})
we can easily calculate the potential $U_s(v,z)$ on $S^*$, where
$v=\varrho \theta$. The Hamiltonian of single particle can be
expressed as $ H'=-\frac{\hbar^2}{2\mu}(\frac{\partial^2}{\partial
v^2}+\frac{\partial^2}{\partial z^2})+U_s(v,z). $

In fact, $U_s$ has periodic structure. If we denote ${\bm
\alpha}_1=(2\pi\varrho/N,\tau)$, ${\bm \alpha}_2=(0,T)$ and ${\bm
\gamma}_l=l_1{\bm \alpha}_1+l_2{\bm \alpha}_2$, we have $U_s({\bf
r}+{\bm \gamma}_l)=U_s({\bf r})$, where $l_1,l_2\in \mathbb{Z}$
and ${\bf r}=(v,z)$. On the one hand, we have the Bloch's theorem
\cite{kittel}:
\begin{equation}\label{plane}
\begin{array}{l}
H'\Phi({\bm \kappa},{\bf r})=E_{\bm \kappa}\Phi({\bm \kappa},{\bf r}),\\
\Phi({\bm \kappa},{\bf
r}+{\bm\gamma}_l)=e^{i{\bm\kappa}\cdot{\bm\gamma}_l}\Phi({\bm\kappa},{\bf
r}),\end{array}
\end{equation}
which suggests that $\Phi({\bm \kappa},{\bf
r})=\sum_ja({\bm\kappa}+{\bf G}_j)e^{i({\bm\kappa}+{\bf G}_j)\cdot
{\bf r}}$, where ${\bf G}_j=j_1{\bm \beta}_1+j_2{\bm \beta}_2$
with $j_1,j_2\in \mathbb{Z}$, ${\bm \beta}_1=(N/\varrho,0)$ and
${\bm \beta}_2=(-\tau N/(T\varrho),2\pi/T)$. From
Eq.(\ref{plane}), we obtain the secular equation
\begin{equation}\label{scu2}
det(\mathcal{H}_{lj}-E_{\bm \kappa}\delta_{lj})=0,
\end{equation}
where $\mathcal{H}_{lj}=\frac{\hbar^2}{2\mu}({\bm\kappa}+{\bf
G}_j)^2\delta_{lj}+\mathcal{U}_{lj}$,
$\mathcal{U}_{lj}=\frac{1}{\Omega_0}\int_{\Omega_0}e^{i({\bf
G}_l-{\bf G}_j)\cdot {\bf r}}U_s({\bf r})d{\bf r}$ and
$\Omega_0=|{\bm \alpha}_1\times{\bm \alpha}_2|$.

On the other hand, Periodic boundary condition along the
circumference of the shell $S^*$ suggests that we just need to
consider the first Brillouin zone which consists of ${\bm
\kappa}=(m_v/\varrho,\kappa_z)$ where $m_v\in\mathbb{Z}, 0\leq
m_v<N$ and $\kappa_z\in\mathbb{R}, 0\leq \kappa_z<2\pi/T$.

From Eq.(\ref{scu2}), we can calculate the energy
$E_{m_v,\kappa_z}$, and then the free energy
$\mathcal{F}=-\mathcal{NT}\ln\mathcal{Z}_1$ \cite{remark1}, where
$\mathcal{Z}_1=\sum_{m_v}\int_0^{2\pi/T}e^{-E_{m_v,\kappa_z}/\mathcal{T}}d\kappa_z$.
Moreover, the specific heat per atom is \cite{landau2}
\begin{equation}
c_v=-\frac{\mathcal{T}\partial^2\mathcal{F}}{\mathcal{N}\partial\mathcal{T}^2}=\frac{\langle
E'^2\rangle-\langle E'\rangle^2}{\mathcal{T}^2},
\end{equation}
where $\langle
E'\rangle=\sum\limits_{m_v=0}^{N-1}\int_0^{2\pi/T}E_{m_v,\kappa_z}e^{-E_{m_v,\kappa_z}/\mathcal{T}}d\kappa_z/\mathcal{Z}$
and $\langle
E'^2\rangle=\sum\limits_{m_v=0}^{N-1}\int_0^{2\pi/T}E^2_{m_v,\kappa_z}e^{-E_{m_v,\kappa_z}/\mathcal{T}}d\kappa_z/\mathcal{Z}$.

In Fig.\ref{fig5}, the symbols ``$\triangledown$" reflect $c_v$
varying with the temperature $\mathcal{T}$, which implies Ne atoms
inside (20,20) tube behave as the lattice gas \cite{cole2} at
ultra low temperature (specific heat is 0) and 2D gas at low
temperature (specific heat approaches 1). There is no 1D gas
inside (20,20) tube, which is quite different from our usual
notion.

\section{\label{classic}why not use CSM?}
At the sufficient high temperature, $\mathcal{F}$ converge to the
$\mathcal{F}_{cl}$. If a carbon nanotube, for example (20, 20)
tube, just a geometric tube with diameter $d\approx 2$ nm, the
characteristic temperature $\mathcal{T}_c=h^2/(2\pi m d^2)=0.04$
K. We just need to use the CSM if $\mathcal{T}>>\mathcal{T}_c$.
However, here the carbon nanotube provides potential to Ne. In the
potential, it is difficult matter to calculate the characteristic
temperature is not a simple matter. For example, consider some
atoms in a harmonic potential $U(r)=\frac{1}{2} m \omega^2 r^2$.
We can easily calculate the specific heat per atom
$c_v=(3\omega^2/\mathcal{T}^2)
\exp(-\omega/\mathcal{T})/[1-\exp(-\omega/\mathcal{T})]^2$
\cite{landau2}. Here we have set $\hbar=1$. We know $c_v=3$ if
$\mathcal{T}>>\omega$ and this is the classical case. Thus we can
set $\mathcal{T}_c=\omega$ (In the whole form $\mathcal{T}_c=\hbar
\omega/k_B$). Above discussion suggests that $\mathcal{T}_c$
depends on the potential. We propose a coarse method to obtain
$\mathcal{T}_c$ which is the least root of the following equation:
\begin{equation}\label{eq1}
\left\{\begin{array}{l}U_m+\mathcal{T}_c=U(R_1)=U(R_2)\\
R_2-R_1=h/\sqrt{2\pi m\mathcal{T}_c}
\end{array}\right..
\end{equation}
The physical meanings of parameters in Eq.(\ref{eq1}) are shown in
the Fig.\ref{rsm1}.

For the potential $U(r)=\frac{1}{2} m \omega^2 r^2$ in above
example, $U_m=0, R_1=-R_2, R_2-R_1=2R_2$, using Eq.(\ref{eq1}) we
arrive at $\mathcal{T}_c=\frac{\sqrt{\pi}}{2}\omega\sim \omega$
(the whole form is
$\mathcal{T}_c=\frac{\sqrt{\pi}\hbar\omega}{2k_B} \sim \hbar
\omega/k_B$). For (20,20) nanotube, the approximate potential is
$U(r)=4\varepsilon[(\frac{\tilde{\sigma}}{\rho-r})^{10}-(\frac{\tilde{\sigma}}{\rho-r})^{4}]$,
with $\rho=13.56$ \AA, $\varepsilon=390$ K, and
$\tilde{\sigma}=2.63$ \AA. We can numerical solve Eq.(\ref{eq1})
and give $\mathcal{T}_c\approx 60$ K. If
$\mathcal{T}>>\mathcal{T}_c$ (e.g.
$\mathcal{T}>3\mathcal{T}_c=180$ K), the classical case is
expected and CSM is valid. But CSM can not be used if
$\mathcal{T}\sim \mathcal{T}_c$ (e.g.
$\mathcal{T}<2\mathcal{T}_c=120$ K), especially
$\mathcal{T}<\mathcal{T}_c=60$ K.

\begin{figure}[!htp]
\begin{center}\includegraphics[width=7cm]{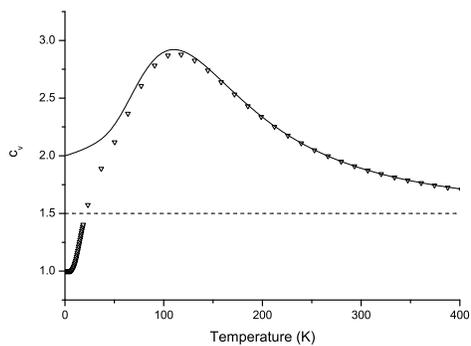}\end{center}
\caption{\label{fig4} The specific heats per atom $c_v$ of Ne
inside $(20,20)$ nanotube at different temperatures without
considering the $\theta,z$ effect. The triangles is the results of
OS and the solid line is that of CSM.}
\end{figure}

\begin{figure}[!htp]
\begin{center}\includegraphics[width=7cm]{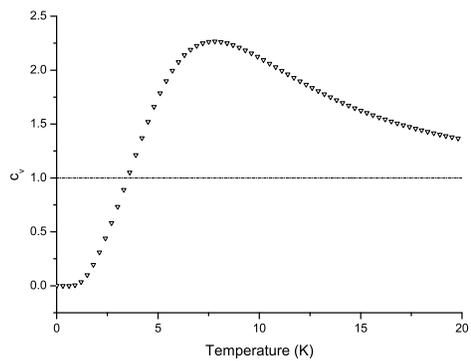}\end{center}
\caption{\label{fig5}The specific heats (triangles) per atom $c_v$
of Ne inside $(20,20)$ nanotubes at low temperatures.}
\end{figure}

\begin{figure}[!htp]
\begin{center}\includegraphics[width=7cm]{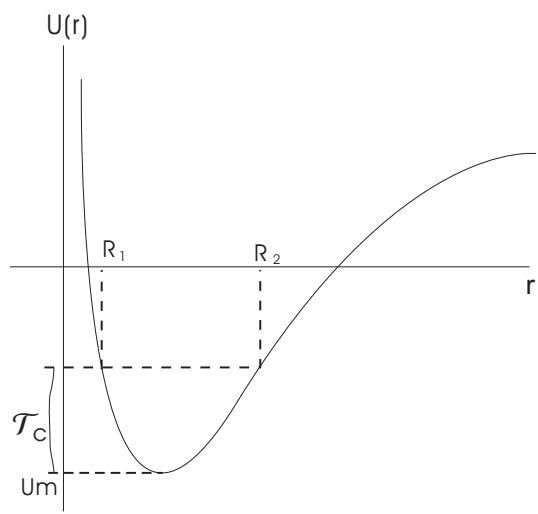}\end{center}
\caption{\label{rsm1} A potential and its characteristic
temperature. $U_m$ is the minimum of the potential and
$\mathcal{T}_c$ is the characteristic temperature. The parameters
$R_1$ and $R_2$ satisfy $U(R_1)=U(R_2)=U_m+\mathcal{T}_c$.}
\end{figure}

In Fig.\ref{fig4} we give results obtained from OS and CSM. The
information shown in Fig.\ref{fig4} agrees with our coarse
estimate to $\mathcal{T}_c$ very well. Therefore, using the CMS is
insufficient. It is necessary to consider the quantum mechanics.
In other words, quantum mechanics must be used when studying the
behaviors of gases in the nanometer scale.

\section{\label{discuss}discussion and conclusion}
\begin{figure}[!htp]
\begin{center}\includegraphics[width=7cm]{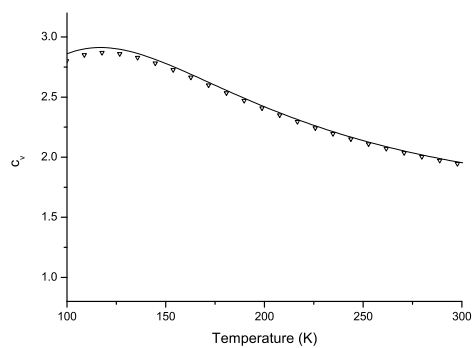}\end{center}
\caption{\label{rsmb1}The specific heats per atom $c_v$ of Ne
inside $(20,20)$ nanotube at high temperatures. The triangles is
the results of OS without considering the $\theta,z$ effect and
the solid line is that of CSM with the $\theta,z$ effect.}
\end{figure}

We notice that the potential $U$ is a function of $(r,\theta,z)$
and $U(r,\theta+2\pi j_1/N,z+j_1\tau+j_2T)=U(r,\theta,z)$, where
$N, \tau, T$ are parameters of a carbon nanotube, and $j_1, j_2$
are two integers. Under this symmetry, we must use a generalized
Bloch's theorem (see appendix) to find the eigenvalues of Schr\"{o}dinger
equation. We can also obtain the secular equation of energies.
Although we have 16
 Pentium IV computers, but the data and
computing time exceed our computers' ability if we calculate the
specific heat with temperature varying from 0 to 400 K. But,
fortunately, we find that the variation of potential with
$\theta,z$ is less than 30 K, much smaller than that with $r$. In
Sec.\ref{cvhigh}, we neglect the effect of the variation of
potential with $\theta,z$. Indeed, this effect can be neglected
under high temperature in terms of intuition. At high temperature,
the values of $c_v$ approach to that calculated from CSM. We show
them in Fig.\ref{rsmb1}, in which the triangles are values
neglected the effect of $\theta,z$ and the line is the result
calculated from CSM considering the effect of $\theta,z$. It
suggests our intuition is reliable.

We know the effect of $\theta,z$ is important under the low
temperature. In Sec.\ref{cvlow}, we consider this effect in
detail. Because the shape of $U$ is very sharp and deep with r,
the particles are confined in the bottom of $U(r)$ at low
temperature. Thus we can consider the effect of $\theta,z$ but
neglect the effect of $r$ at low temperature.

Therefore, we believe that we have grasped the main factors under
at high and low temperature and our qualitative conclusions are
reliable.

The interactions (include the contribution from the polarization
of the pi-electrons) between Ne and C atoms are intermolecular
forces, van der Waals type interactions. As the theoretical study,
the simplest form of van der Waals force is the Lenard-Jones
potential. We believe that the eventual results based on the
Lenard-Jones potential are qualitatively reasonable. Maybe, many
researchers \cite{cole,carraro} use it because of the same
reasons. Indeed, the more detailed study needs to consider the
Gay-Berne potential \cite{Gay} or other complex potentials. We are
considering to overcome this difficult problem by using numerical
simulations.

It is necessary to notice that we just consider the dilute Ne and
the interactions between Ne atoms are naturally neglected.
Therefore we do not describe the phase transition of Ne in SWNT's.
If we intend to deal with the phase transition, we must include
the interactions between Ne atoms and use the methods of quantum
field theory, such as the renormalization group approach
\cite{renormal}, etc.

In conclusion, we calculate the specific heat of Ne in (20,20)
SWNT at low and high temperature and find the dimensional
crossover of thermal behavior in this system. Especially,
the dilute Ne gas exhibits as 2D behavior at low temperature
or 2D lattice behavior at ultra low temperature. The simple physical picture is that all atoms are confined on a shell blow a critical temperature.

\ack{ We are grateful to kind discussion of Prof. H. W. Peng, Dr.
R. An, M. Li, Y. Zhang, and L. R. Dai.}

\section*{Appendix: Generalized Bloch's theorem}
In Sec.\ref{cvhigh}, we calculate the specific heat with
neglecting the effect of $\theta,z$; and in Sec.\ref{cvlow}, in
fact, we neglect the effect of $r$. In other words, we do not
obtain the specific heats in a consistent way from low to high
temperature. Indeed, the potential $U$ is a function of
$(r,\theta,z)$ and satisfies $U(r,\theta+2\pi
j_1/N,z+j_1\tau+j_2T)=U(r,\theta,z)$, where $j_1, j_2$ are two
integers. If a potential satisfies this condition, it is called
the spiral symmetric. Under this spiral symmetry, we must use a
generalized Bloch's theorem to find the eigenvalues of
Schr\"{o}dinger equation and then calculate the specific heat in a
consistent way \cite{tzc2}.

Above all, we go over traditional translational symmetry and
Bloch's theorem \cite{Callaway}.

The hamiltonian of a system with the translational symmetry is
expressed as
\begin{equation}H=-\frac{\hbar^2}{2\mu} \nabla^2+V({\bf r});\quad  V({\bf r}+{\bf R}_j)=V({\bf r}),\end{equation}
where ${\bf R}_j=n_{j1}{\bf a}_1+n_{j2}{\bf a}_2+n_{j3}{\bf a}_3$
and $n_{j1},n_{j2},n_{j3}$ are integers.

Define translational operators $\mathcal{J}({\bf R}_j)$, which act
on a function $f({\bf r})$ as:
\begin{equation}\mathcal{J}({\bf R}_j)f({\bf r})=f({\bf r}+{\bf
R}_j).\end{equation} It follows that
\begin{equation}\label{b4}\mathcal{J}({\bf R}_j)\mathcal{J}({\bf R}_l)=\mathcal{J}({\bf
R}_l+{\bf R}_j)=\mathcal{J}({\bf R}_j+{\bf R}_l)=\mathcal{J}({\bf
R}_l)\mathcal{J}({\bf R}_j),\end{equation} and
\begin{eqnarray}\mathcal{J}({\bf R}_j)Hf({\bf r})=-\frac{\hbar^2}{2\mu}
\mathcal{J}({\bf R}_j)\nabla^2f({\bf r})+\mathcal{J}({\bf
R}_j)V({\bf r})f({\bf r})\nonumber\\
 =-\frac{\hbar^2}{2\mu}\nabla^2 \mathcal{J}({\bf
R}_j)f({\bf r})+V({\bf r}+{\bf R}_j)f({\bf r}+{\bf R}_j)\nonumber \\
=[-\frac{\hbar^2}{2\mu}\nabla^2+V({\bf r})]f({\bf r}+{\bf
R}_j)=H\mathcal{J}({\bf R}_j)f({\bf r}).\end{eqnarray} Therefore
$\{\mathcal{J}({\bf R}_j),H\}$ is the set of conserved quantities.
In this case, an eigenfunction of the Hamiltonian must be an
eigenfunction of the translational operators, i.e.,
$\mathcal{J}({\bf R}_j)\psi({\bf r})=\psi({\bf r}+{\bf
R}_j)=\lambda({\bf R}_j)\psi({\bf r})$ if $H\psi({\bf
r})=E\psi({\bf r})$. Furthermore, the electron density must be
periodic, i.e., $|\psi({\bf r}+{\bf R}_j)|^2=|\psi({\bf r})|^2$.
It follows that
\begin{equation}\label{b8}|\lambda({\bf R}_j)|^2=1.\end{equation}
But from Eq.(\ref{b4}), we know
\begin{equation}\label{b7}\lambda({\bf R}_j)\lambda({\bf R}_l)=\lambda({\bf R}_j+{\bf R}_l).\end{equation}
The solution of Eq.(\ref{b7}) under the constraint Eq.(\ref{b8})
is $\lambda({\bf R}_j)=e^{i{\bm \kappa}\cdot {\bf R}_j}$. Thus we
have Bloch's theorem:
\begin{equation}\label{b10}\psi({\bf r}+{\bf
R}_j)=e^{i{\bm \kappa}\cdot {\bf R}_j}\psi({\bf r}).\end{equation}

Next, we will turn to the spiral symmetry and generalize the
Bloch's theorem. The hamiltonian with the spiral symmetry in
cylindrical coordinates is expressed:
\begin{equation}\label{hspiral}H=-\frac{\hbar^2}{2\mu} (\frac{\partial^2}{\partial r^2}+\frac{\partial}{r\partial r}+\frac{\partial^2}{r^2\partial\theta^2}+\frac{\partial^2}{\partial z^2})+V(r,\theta,z),\end{equation}
\begin{equation}\label{spiral}V(r,\theta,z)=V(r,\theta+j_1\vartheta,z+j_1\tau+j_2T),\end{equation}
where $\vartheta=2\pi/N, N\tau=MT$, $N,M\in\mathbb{N}$,
$j_1,j_2\in \mathbb{Z}$. Define vectors ${\bf r}=(\theta, z)$,
${\bf R}_j=(j_1\vartheta,j_1\tau+j_2T)$ in the space
$[0,2\pi)\times \mathbb{R}$, and operators $\mathcal{J}({\bf
R}_j)$ which act on a function $f(r;{\bf r})$ as $\mathcal{J}({\bf
R}_j)f(r;{\bf r})=f(r;{\bf r}+{\bf R}_j)$. It follows that
$\mathcal{J}({\bf R}_j)\mathcal{J}({\bf R}_l)=\mathcal{J}({\bf
R}_l+{\bf R}_j)=\mathcal{J}({\bf R}_j+{\bf R}_l)=\mathcal{J}({\bf
R}_l)\mathcal{J}({\bf R}_j)$ and $\mathcal{J}({\bf R}_j)V(r;{\bf
r})=V(r;{\bf r})\mathcal{J}({\bf R}_j)$.
Otherwise,\begin{equation}\mathcal{J}({\bf
R}_j)(\frac{\partial^2}{r^2\partial\theta^2}+\frac{\partial^2}{\partial
z^2})f(r;{\bf
r})=(\frac{\partial^2}{r^2\partial\theta^2}+\frac{\partial^2}{\partial
z^2})\mathcal{J}({\bf R}_j)f(r;{\bf r}).\end{equation} Thus
\begin{equation}\label{bb3}\mathcal{J}({\bf R}_j)H=H\mathcal{J}({\bf
R}_j).\end{equation} We can obtain generalized Bloch's theorem
analogizing the method to obtain the traditional Bloch's theorem:
\begin{equation}\label{gbloch}
\psi(r;{\bf r}+{\bf R}_j)=e^{i{\bm \kappa}\cdot {\bf
R}_j}\psi(r;{\bf r}),\end{equation} where ${\bm
\kappa}=(l,\kappa)$, $l=0,1,\cdots,N-1,\quad 0\leq \kappa<2\pi/T$.

If we set ${\bm \alpha}_1=(\vartheta, \tau)$, ${\bm
\alpha}_2=(0,T)$, ${\bm \beta}_1=(N,0)$, ${\bm \beta}_2=(-\tau
N/T,2\pi/T)$, ${\bf G_j}=j_1{\bm \beta}_1+j_2{\bm
\beta}_2=(j_1N-j_2N\tau/T, 2\pi j_2/T)$, then we can expand the
wave function as the superposition of planar waves:
\begin{equation}\label{plane}\psi(r;{\bf
r})=\sum\limits_{{\bf j}}C_{{\bf j}}\varphi(r)e^{i({\bm
\kappa}+{\bf G_j})\cdot {\bf r}}.\end{equation}

In cylindrical coordinates, Eqs.(\ref{gbloch}) and (\ref{plane})
are expressed
\begin{eqnarray}\label{blc1}
\psi(r,\theta+\frac{2\pi j_1}{N},z+j_1\tau+j_2T)=e^{i[\frac{2\pi
lj_1}{N}+\kappa (j_1\tau+j_2T)]}\psi(r,\theta,z),\\
\label{plane2}\psi_{l\kappa}(r,\theta,z)=\sum\limits_{j_1j_2}C_{j_1j_2}\varphi_{l\kappa}(r)e^{i[(l+j_1N-j_2N\tau/T)\theta+(\kappa+2\pi
j_2/T)z]}.
\end{eqnarray}

If Expanding $\varphi_{l\kappa}(r)$ with normalized orthogonal
basis $\{\chi_{nl\kappa}(r), n\in \mathbb{N}\}$ we obtain the
secular equation from Eq.(\ref{plane2}) and Shr\"{o}dinger
equation
\begin{equation}det[(\mathcal{H}_{m_1m_2j_1j_2nn'}-E_{nl\kappa}\delta_{j_1m_1}\delta_{j_2m_2}\delta_{nn'})]=0,\end{equation}
where
$\mathcal{H}_{j_1j_2m_1m_2nn'}=\mathcal{T}_{j_1j_2nn'}\delta_{j_1m_1}\delta_{j_2m_2}+\mathcal{U}_{j_1j_2m_1m_2nn'}$,
$\mathcal{T}_{j_1j_2nn'}=-\frac{\hbar^2}{2\mu}\int_0^{\rho}\chi_{nl\kappa}(r)[\frac{d^2}{dr^2}+\frac{d}{r
dr}-\frac{(l+j_1N-\frac{j_2N\tau}{T})^2}{r^2}-(\kappa+\frac{2\pi
j_2}{T})^2]\chi_{n'l\kappa}(r)rdr$, and\\
$\mathcal{U}_{j_1j_2m_1m_2nn'}= \frac{N}{2\pi
T}\int_0^{2\pi}d\theta\int_0^Tdz\int_0^{\rho}rdr\chi_{nl\kappa}(r)U(r,\theta,z)\chi_{n'l\kappa}(r)e^{i[(j_1-m_1)-\frac{(j_2-m_2)\tau}{T}]N\theta+\frac{i2\pi
(j_2-m_2)z}{T}}$.

\newpage

\end{document}